\pdfoutput=1
\documentclass[journal]{IEEEtran}
\usepackage{amsmath,amssymb,amsthm}
\usepackage{graphicx}
\usepackage{booktabs}
\usepackage{cite}
\usepackage{hyperref}

\newtheorem{lemma}{Lemma}
\newtheorem{proposition}{Proposition}

\theoremstyle{remark}
\newtheorem{remark}{Remark}

\newcommand{\E}{\mathbb{E}}
\newcommand{\Rea}{\operatorname{Re}}
\newcommand{\diag}{\operatorname{diag}}

\begin{document}

\title{Which Chains to Calibrate for Physical Backpropagation:\\ Per-User Factors Versus Base-Station Chains}

\author{Xianzhong~Li, Yuzhi~Yang, and Tianhao~Guo
\thanks{Xianzhong Li and Tianhao Guo are with the College of Physics and Electronic Engineering, Shanxi University, Taiyuan 030006, China (e-mail: \{lixianzhong, tianhao\_guo\}@sxu.edu.cn).}%
\thanks{Yuzhi Yang is with the College of Information Science, Japan Advanced Institute of Science and Technology, Nomi 9231292, Japan (e-mail: yangyzis@jaist.ac.jp).}%
\thanks{Corresponding author: Tianhao Guo.}}

\maketitle

\begin{abstract}
In physical backpropagation, users return loss derivatives over the time-division duplex uplink, giving the base station the downlink gradient without channel reconstruction. Reciprocity calibration conventionally prioritizes base-station chains, as column-normalized zero-forcing reduces a per-user factor to a stream rotation. This priority can reverse. With ideal base-station chains and no noise, per-user amplitude errors make the returned field a weighted sum-rate gradient. With zero-forcing full service at both optima, the loss is independent of array size. Otherwise the selection penalty, serving the wrong users, carries 86 to 94 percent of the amplitude-induced loss in five of six configurations. Per-user phase errors mainly slow the iteration. Base-station chain errors scale the returned field per antenna, and at 3 dB and 25 degrees they cost one user 1.3 dB of signal-to-noise ratio (SNR) that no array size removes. With 80 users, such errors on both sides, and mismatched pilots, oracle per-user and base-station calibration leave 6.2 and 12.3 percent loss at 256 antennas on clustered channels. This ordering holds at 30 dB SNR and reverses for 1 dB, 10 degree user errors. A capped one-echo estimator comes within 2.4 percentage points of oracle per-user calibration but underperforms no calibration at 10 dB SNR.
\end{abstract}


\section{Introduction}

When the same physical path carries a signal forward and an error signal backward, the backward pass can supply the adjoint of the forward operator that gradient methods need. Optical and wave-based learning hardware exploits this property \cite{hughes2018optica,wright2022nature}, and recent work states general conditions under which a physical system can perform its own backpropagation \cite{bosch2026unifying,niu2026reciprocity}. In time-division duplex (TDD) systems, the uplink propagation channel equals the transpose of the downlink propagation channel \cite{marzetta2010noncooperative}, and with reciprocal hardware and a conjugation at each end the uplink returns the complex adjoint. Over-the-air split learning \cite{yang2023overtheair} and over-the-air gradient methods for precoding \cite{dai2025quasi,yang2026p2gd} use this to let the users return their loss derivatives over the uplink, so that the base station (BS) receives the gradient of the downlink objective without reconstructing the channel. In all of these systems the users return the derivatives of their own losses over the uplink, and Lemma~\ref{lem:decomp}, Proposition~\ref{prop:bs}, and the first claim of Proposition~\ref{prop:ue} hold for any of them, since they use only that the objective is a sum of per-user terms. We use downlink precoding to compute what each error costs.

Our companion papers study how channel aging breaks this adjoint in time and how waiting, rotating, or moving the antennas repairs it \cite{li2026wait,li2026rotate,li2026pace}. This paper studies the hardware. The baseband channel includes the transmit and receive radio-frequency chains on both ends, and their gains differ \cite{bourdoux2003nonreciprocal}. Reciprocity calibration is a mature subject \cite{guillaud2005practical,shepard2012argos,rogalin2014scalable,vieira2017reciprocity,jiang2018framework}, and its practice gives priority to the BS side \cite{jiang2018framework}. A per-user factor multiplies one column of the effective channel. A zero-forcing precoder then carries the inverse of its conjugate on that user's column. Per-column normalization removes the amplitude, and the user's receiver removes the phase with downlink pilots. With a global normalization the factor changes the power allocation instead. User-side errors still matter in some settings. Without downlink demodulation pilots they degrade the rate \cite{raeesi2018performance,raeesi2018estimation}, and in cell-free systems the rate is more sensitive to user-side calibration error \cite{fang2025cellfree}. The gradient-based precoding works cited above assume ideal reciprocity, and \cite{yang2026p2gd} names chain mismatch as an open issue. In optical and wave hardware, a known diagonal mismatch can be compensated to recover the exact adjoint \cite{li2026symipac,bosch2026unifying}. Here the factors are unknown. Over-the-air calibration frameworks already involve the users \cite{jiang2018framework}, and the BS and user non-reciprocity can be estimated jointly \cite{raeesi2018estimation}. These works serve a linear precoder computed from the calibrated channel. In physical backpropagation the precoder is computed from the gradient that the users return, and this changes which factors matter.

In physical backpropagation, a per-user factor multiplies that user's contribution to the returned field, much as heterogeneous client weights change the objective that federated optimization solves \cite{wang2020objective}. This paper asks which chains a physical backpropagation system must calibrate, and how the answer changes with the array size. Our contributions are as follows.

\begin{itemize}
\item We decompose the returned field into a BS factor and a per-user factor (Lemma~\ref{lem:decomp}). With ideal BS chains and noiseless forward outputs and returns, the per-user amplitude errors make the returned field the gradient of a weighted sum rate (Proposition~\ref{prop:ue}). When the beams are restricted to zero-forcing directions and both the plain and the weighted optima serve every user, the sum-rate loss equals the number of users times the base-two logarithm of the ratio between the arithmetic and geometric means of the weights, independent of the channel gains and the array size (Proposition~\ref{prop:fullservice}). We split the general loss into a selection penalty and an allocation penalty (Proposition~\ref{prop:split}) and show that the selection penalty can be arbitrarily large (Proposition~\ref{prop:group}).
\item In simulations with 80 users, the selection penalty carries 87 to 93\% of the amplitude-induced loss in the five configurations where the endpoints leave many users unserved. Keeping the better of the iteration and an independent weighted minimum mean square error (WMMSE) solver in every optimization of the split gives $86$ to $94$\%. On an exactly solvable instance the solver protocol understates the selection penalty rather than inflating it. Per-user phase errors cost about $24$\,bit/s/Hz after 3000 iterations at 256 antennas, but most of this disappears by $10\,000$ iterations, and on the Rayleigh channel the remainder matches the cosine-weight estimate.
\item The BS chains scale the returned field diagonally. With ideal per-user factors and noiseless forward outputs and returns, the scaled field remains an ascent direction of the unconstrained objective when every chain has a phase error below $90^\circ$, and its alignment concentrates around a constant set by the error statistics (Proposition~\ref{prop:bs}). For one user every returned field is a multiple of one beam, whose signal-to-noise ratio (SNR) loss has a closed form that a larger array does not remove, and in simulation the iteration reaches this beam (Proposition~\ref{prop:single}).
\item With pilots that pass through the mismatched chains and paired hardware draws, oracle per-user calibration beats BS calibration on clustered channels at every tested array size, and on Rayleigh channels up to 128 antennas. Against BS calibration relative to one antenna it leads at every tested array size on both channels. We map where the ordering reverses. We give an estimator that obtains each per-user factor from one uplink echo and one reported scalar (Lemma~\ref{lem:e1}) and state when it helps.
\end{itemize}

Section~\ref{sec:model} gives the model. Section~\ref{sec:analysis} analyzes the two sides. Section~\ref{sec:e1} gives the estimator. Section~\ref{sec:sim} reports the simulations, and Section~\ref{sec:concl} concludes.

\section{System Model}\label{sec:model}

A BS with $M$ antennas serves $K$ single-antenna users. The downlink channel is $H=[h_1,\dots,h_K]\in\mathbb C^{M\times K}$ and the precoder is $W=[w_1,\dots,w_K]$ with $\|W\|_F^2=P$. User $k$ observes the outputs $y_{kj}=h_k^Hw_j$ of a forward probe and has the rate
\begin{equation}
R_k(W)=\log_2\Bigl(1+\frac{|y_{kk}|^2}{\sum_{j\ne k}|y_{kj}|^2+\sigma^2}\Bigr).
\end{equation}
Here $\sigma^2$ is the receiver noise variance. The objective is the sum rate $R=\sum_kR_k$. Rates and rate differences are reported in bit/s/Hz.

\begin{figure*}[t]
\centering
\includegraphics[width=\textwidth]{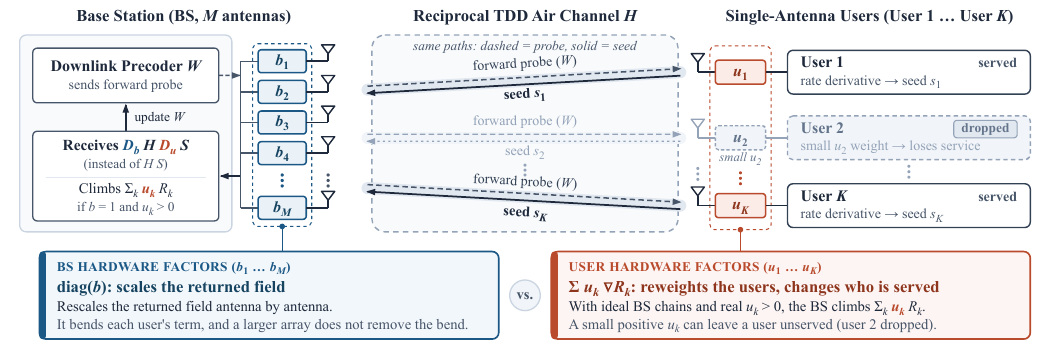}
\caption{Physical backpropagation with non-reciprocal hardware. The BS sends a forward probe with $W$ (dashed), each user returns its seed, the derivative of its rate, over the same paths (solid), and the BS updates $W$ from what it receives. The BS chain factors $b_i$ scale the returned field antenna by antenna. The user factors $u_k$ weight each user's term. With ideal BS chains, no noise, and positive real $u_k$ the BS climbs a weighted sum rate, and a user with a small weight can lose service.}
\label{fig:system}
\end{figure*}

\subsection{Physical Backpropagation}
Fig.~\ref{fig:system} shows the loop. User $k$ computes its seed $s_k^\top=\partial(R_k\ln2)/\partial\bar y_{k,:}$, the derivative of its rate in nats, which depends only on its own outputs. The downlink operator is $H^H$, so a reciprocal uplink applies $(H^H)^\top=\bar H$. User $k$ therefore transmits $\bar s_k$, and the BS conjugates what it receives. With ideal reciprocity and no noise the BS obtains
\begin{equation}
G=\sum_k h_k s_k^\top=\ln2\,\nabla_{\bar W}R.
\end{equation}
The positive factor $\ln2$ is removed by normalization. At step $t$ the BS forms the momentum $v_t=0.9\,v_{t-1}+0.1\,\hat G_t/\|\hat G_t\|_F$ from the returned field $\hat G_t$ and sets $W_{t+1}=\Pi(W_t+2\cdot0.999^t\,v_t)$, where $\Pi$ rescales onto the power sphere. The iterations do not reconstruct the channel. In our experiments the iteration starts from a pilot-based initial precoder, described in Section~\ref{sec:sim}.

\subsection{Hardware Non-Reciprocity}
The baseband uplink includes the BS receive chains and the user transmit chains, and the baseband downlink includes the BS transmit chains and the user receive chains. Only their ratios matter for the adjoint \cite{guillaud2005practical}. We collect the ratios that remain after the two conjugations into a BS factor $b\in\mathbb C^M$ and a per-user factor $u\in\mathbb C^K$, so that the returned field is
\begin{equation}\label{eq:ghat}
\hat G=D_b\,H\,D_u\,S,\qquad D_b=\diag(b),\ D_u=\diag(u),
\end{equation}
plus receiver noise, where $S\in\mathbb C^{K\times K}$ stacks the seeds $s_k^\top$ as rows. Each factor has independent lognormal amplitude and Gaussian phase, $b_i=10^{\alpha_i/20}e^{j\theta_i}$ with $\alpha_i\sim\mathcal N(0,\sigma_{\rm dB}^2)$ and $\theta_i\sim\mathcal N(0,\sigma_\theta^2)$, and the same, with independent draws, for $u_k$. All draws are independent across antennas and users, and $\sigma_\theta$ is in radians in all formulas. The mismatch is static over the optimization. We write $\sigma_{\ln}=\sigma_{\rm dB}\ln 10/20$.

\section{Where Each Side Enters}\label{sec:analysis}

\begin{lemma}[Two-sided decomposition]\label{lem:decomp}
Without noise, $\hat G=\ln2\,D_b\sum_k u_k\nabla_{\bar W}R_k$.
\end{lemma}
\begin{proof}
Substitute $S$ row by row into \eqref{eq:ghat} and use $h_ks_k^\top=\ln2\,\nabla_{\bar W}R_k$.
\end{proof}

The decomposition holds for any loss that is a sum of per-user terms.

The BS factor multiplies the whole gradient from the left. The per-user factor multiplies the contribution of one user to the objective, so the two sides act on different objects.

\subsection{The BS Side Scales the Gradient}

\begin{proposition}[BS side]\label{prop:bs}
Let $u\equiv1$, let the forward outputs and the returns be noiseless, let $G\neq0$, and let $g_i$ denote the $i$th row of $G$. We write $\langle A,B\rangle=\operatorname{tr}(A^HB)$ and $\cos\angle(A,B)=\Rea\langle A,B\rangle/(\|A\|_F\|B\|_F)$.
\begin{enumerate}
\item $\Rea\langle G,\hat G\rangle=\sum_{i=1}^M\Rea(b_i)\|g_i\|^2$. If $\Rea\,b_i>0$ for every $i$, then $\hat G$ is an ascent direction of $R$ at $W$ when $W$ is unconstrained.
\item Let $(H,W)$ be fixed and independent of $b$, and let the $b_i$ be independent and identically distributed with finite fourth moment and $\E|b|^2>0$. Let $\omega_i=\|g_i\|^2/\|G\|_F^2$ and $N_{\rm eff}=1/\sum_i\omega_i^2$. Then $\cos\angle(G,\hat G)$ converges in probability to $\E\Rea\,b/\sqrt{\E|b|^2}$ as $N_{\rm eff}\to\infty$, with fluctuations of order $N_{\rm eff}^{-1/2}$.
\end{enumerate}
\end{proposition}
\begin{proof}
The first identity is a row-by-row expansion of the inner product. For the second, write the cosine as $\sum_i\omega_i\Rea\,b_i\big/\bigl(\sum_i\omega_i|b_i|^2\bigr)^{1/2}$. The numerator and the squared denominator are weighted averages of independent variables with weights summing to one, so each has variance equal to its single-term variance divided by $N_{\rm eff}$, which is finite under the fourth-moment assumption, and Chebyshev's inequality gives their convergence to $\E\Rea\,b$ and $\E|b|^2>0$ with deviations of order $N_{\rm eff}^{-1/2}$. The map $(x,y)\mapsto x/\sqrt y$ is continuously differentiable at this limit, so the cosine converges with deviations of the same order.
\end{proof}

For $3$\,dB and $25^\circ$ errors the limit is $e^{-(\sigma_{\ln}^2+\sigma_\theta^2)/2}\approx0.86$, an angle of about $31^\circ$. The array shrinks the fluctuation around this angle but not the angle itself. A single chain violates $\Rea\,b_i>0$ with probability about $2Q(\pi/(2\sigma_\theta))\approx3\times10^{-4}$, where $Q$ is the standard normal tail function, so at least one of $256$ chains violates it with probability about $7.8$\%. The condition is sufficient, not necessary. The power constraint changes the picture. Only the component of $\hat G$ tangent to the power sphere moves $W$, and a non-scalar diagonal scaling can move the fixed points of a projected iteration \cite{bertsekas1982projected}. The BS factor can then also change who is served. With $H=I_2$, diagonal $W$, $P=\sigma^2=1$, and $b=(4,1)$, the returned field is the gradient of $4\ln(1+p_1)+\ln(1+p_2)$, and the iteration ends serving only the first user. Here each user occupies one antenna. In a rich channel write the term of user $k$ as $T_k=h_ks_k^\top$, and let $\mu_k=\sum_i\nu_{ik}b_i$ with $\nu_{ik}=|h_{ik}|^2/\|h_k\|^2$. Then $D_bT_k=\mu_kT_k+(D_bh_k-\mu_kh_k)s_k^\top$, and the second term is orthogonal to $T_k$. The array concentrates the coefficient $\mu_k$ around $\E b$. It does not shrink the orthogonal term. For channels of equal modulus across the antennas, the norm of this term relative to $\|T_k\|_F$ tends to $(\E|b|^2-|\E b|^2)^{1/2}$ as $M$ grows. This is about $0.58$ for $3$\,dB and $25^\circ$ errors, against $|\E b|\approx0.97$. This orthogonal term bends each user's contribution and produces the nonzero limiting angle of Proposition~\ref{prop:bs}, and the power constraint decides how far the bend moves the fixed point. A per-user factor leaves no such term. It multiplies $T_k$ by one number that no array averages. With one user the returned direction has a closed form.

\begin{proposition}[Single-user returned beam]\label{prop:single}
Let $K=1$, $u=1$, and let the forward output and the return be noiseless. Every returned field is a complex multiple of $D_bh$. On the beams $w\propto D_bh$, the SNR relative to the matched filter is
\begin{equation}
\rho=\frac{|\sum_ib_i|h_i|^2|^2}{\|h\|^2\sum_i|b_i|^2|h_i|^2}\le1.
\end{equation}
If the $b_i$ are independent and identically distributed with finite fourth moment, independent of $h$, and $\E|b|^2>0$, then $\rho\to|\E b|^2/\E|b|^2$ in probability as $\|h\|^4/\sum_i|h_i|^4\to\infty$.
\end{proposition}
\begin{proof}
The seed of a single user is a scalar, so $\hat G=D_bh\,s$. The ratio follows from $|h^HD_bh|^2/(\|h\|^2\|D_bh\|^2)$, and the bound from the Cauchy--Schwarz inequality. The limit follows as in Proposition~\ref{prop:bs} with weights $|h_i|^2/\|h\|^2$.
\end{proof}

In our simulations with $M=128$ the iteration of Section~\ref{sec:model} converges to $w\propto D_bh$ and attains this $\rho$ to four digits, from the pilot-based start with $P=K$. This is an observation, not a convergence guarantee. The step sizes are summable, $\sum_t2\cdot0.999^t=2000$, so from a start far enough from this ray relative to the power scale the iteration cannot reach it. For $3$\,dB and $25^\circ$ errors the limit is $0.73$, a loss of about $1.3$\,dB that a larger array does not remove. For one user the BS errors therefore cost a fixed fraction of the SNR rather than a vanishing one. For $K$ users the same fraction per user would suggest a loss of about $K\log_2(\E|b|^2/|\E b|^2)$ bit/s/Hz at high SNR. Section~\ref{sec:sim} finds the multiuser loss below this value.

\subsection{The User Side Changes the Objective}

\begin{proposition}[Per-user amplitude]\label{prop:ue}
Let $b\equiv1$, let the forward outputs and the returns be noiseless, and let $u=a$ with $a_k>0$. Then $\hat G=\ln2\,\nabla_{\bar W}R_a$ with $R_a=\sum_ka_kR_k$, so the returned field is the gradient of a weighted sum rate. If $W_1$ and $W_a$ globally maximize $R$ and $R_a$ on the power sphere, the loss $L=R(W_1)-R(W_a)$ satisfies $0\le L\le(1-a_{\min}/a_{\max})R(W_1)$, where $a_{\min}=\min_ka_k$ and $a_{\max}=\max_ka_k$.
\end{proposition}
\begin{proof}
The first claim is Lemma~\ref{lem:decomp} with $D_b=I$. For the bound, $a_{\min}R(W_1)\le R_a(W_1)\le R_a(W_a)\le a_{\max}R(W_a)$.
\end{proof}

The weights are $K$ fixed numbers, one per user, so the array size does not average them out. What they cost depends on the structure of the optimum. When every user is served with its own spatial dimension, the cost is a reallocation of power.

\begin{proposition}[Full service]\label{prop:fullservice}
Let $M\ge K$ and let $H$ have full column rank. Restrict $W$ to zero-forcing directions with per-user powers, so that user $k$ has signal-to-interference-plus-noise ratio (SINR) $p_k\gamma_k$ with $\gamma_k=1/(\sigma^2[(H^HH)^{-1}]_{kk})$. If every user is active in both the plain and the weighted water-filling solutions, then
\begin{equation}
L=K\log_2\frac{A(a)}{G(a)},
\end{equation}
where $A$ and $G$ are the arithmetic and geometric means of the weights. For independent lognormal weights, $\log_2(A(a)/G(a))\to\sigma_{\ln}^2/(2\ln2)$ almost surely as $K\to\infty$.
\end{proposition}
\begin{proof}
With $c=P+\sum_k1/\gamma_k$, the interior water-filling solutions \cite{he2013waterfilling} are $p_k+1/\gamma_k=c/K$ for the plain objective and $p_k+1/\gamma_k=c\,a_k/(KA)$ for the weighted one. Hence each user's rate changes by exactly $\log_2(a_k/A)$, and summing gives the expression. The limit follows from the strong law of large numbers applied to $A$ and to $\log G$.
\end{proof}

Under these assumptions the loss in Proposition~\ref{prop:fullservice} does not involve $H$, $P$, or $M$, and it holds at any SNR. Full activity under the weighted objective requires $c\,a_k/(KA)>1/\gamma_k$ for every $k$, so it needs per-user SNR that grows with the spread of the weights. With $3$\,dB errors its large-$K$ value is about $6.9$\,bit/s/Hz for $K=80$. As the sum rate grows with the array, its relative size vanishes.

When the optimum serves only part of the users, a second mechanism appears. Let $F(\mathcal S)$ be the maximum sum rate on the power sphere with $w_k=0$ for every $k\notin\mathcal S$. The served set of $W$ is the set of users with $R_k(W)>0$. Let $\mathcal S_1$ and $\mathcal S_a$ be the served sets at $W_1$ and $W_a$.

\begin{proposition}[Selection and allocation]\label{prop:split}
$L=\bigl[F(\mathcal S_1)-F(\mathcal S_a)\bigr]+\bigl[F(\mathcal S_a)-R(W_a)\bigr]$, and both brackets are nonnegative.
\end{proposition}
\begin{proof}
$F(\mathcal S_1)=R(W_1)$ by global optimality of $W_1$, and $F(\mathcal S_1)\ge F(\mathcal S_a)$ by the same. $W_a$ serves $\mathcal S_a$, so $F(\mathcal S_a)\ge R(W_a)$.
\end{proof}

We call the first bracket the selection penalty and the second the allocation penalty. Proposition~\ref{prop:fullservice} gives the second exactly under zero-forcing full service. The first has no such bound.

\begin{proposition}[Large selection penalties exist]\label{prop:group}
Consider $N$ orthogonal spatial groups, each containing two users with collinear channels of gains $g_{g1},g_{g2}$ and a fixed group power $q_g$. Let $c_g>d_g$ be the rates the two users of group $g$ obtain when served alone, and break ties $a_{g2}d_g=a_{g1}c_g$ in favor of the first user. Then every optimum, weighted or not, serves exactly one user per group, and
\begin{equation}
L=\sum_g(c_g-d_g)\,\mathbf 1\{a_{g2}d_g>a_{g1}c_g\}.
\end{equation}
For independent lognormal weights, ties have probability zero and $\E L=\sum_g(c_g-d_g)\,Q\bigl(\ln(c_g/d_g)/(\sqrt2\sigma_{\ln})\bigr)$.
\end{proposition}
\begin{proof}
The groups do not interact, so the objective separates. In group $g$ the two users see the same total power $q_g$, so with $p$ the power of the first user and natural logarithms the weighted objective is
\begin{equation*}
f(p)=a_{g1}\ln\frac{q_gg_{g1}+\sigma^2}{(q_g-p)g_{g1}+\sigma^2}+a_{g2}\ln\frac{q_gg_{g2}+\sigma^2}{p\,g_{g2}+\sigma^2},
\end{equation*}
with $f''(p)=a_{g1}g_{g1}^2/((q_g-p)g_{g1}+\sigma^2)^2+a_{g2}g_{g2}^2/(p\,g_{g2}+\sigma^2)^2>0$. The maximum therefore lies at an endpoint, with values $a_{g1}c_g\ln2$ at $p=q_g$ and $a_{g2}d_g\ln2$ at $p=0$, since $c_g$ and $d_g$ are in bit/s/Hz. The weighted optimizer picks the second user exactly when $a_{g2}d_g>a_{g1}c_g$. The log-ratio of two independent weights is Gaussian with variance $2\sigma_{\ln}^2$.
\end{proof}

The case $N=1$ with $q_1=P$ is the unrestricted single-group problem, so large losses do not rely on the group-power restriction. With $N=40$ groups, $c=10$, and $d=8$\,bit/s/Hz, $3$\,dB errors give an expected loss of about $26$\,bit/s/Hz, nearly four times the full-service value. Scaling $c$ and $d$ together across such channels makes it arbitrarily large. The number of unserved users stays at $N$ throughout, so counting inactive users cannot detect this penalty.

\begin{remark}[Per-user phase]
For $u_k=e^{j\varphi_k}$, the returned field $\ln2\sum_ke^{j\varphi_k}\nabla_{\bar W}R_k$ is generically not the gradient of any real-valued function of $W$, and
$\Rea\langle G,\hat G\rangle=(\ln2)^2\sum_l\Rea\bigl(e^{j\varphi_l}\langle\nabla_{\bar W}R,\nabla_{\bar W}R_l\rangle\bigr)$.
In a power-only update along fixed zero-forcing directions, the per-user factors act exactly as weights $\cos\varphi_l$. If both the plain and the cosine-weighted solutions serve every user and every $\cos\varphi_l>0$, which for 80 users at $25^\circ$ fails with probability $2.5$\%, Proposition~\ref{prop:fullservice} with $\ln\cos\varphi\approx-\varphi^2/2-\varphi^4/12$ gives $\log_2(A/G)\approx\sigma_\theta^4/(4\ln2)$ per user and the small-angle estimate $K\sigma_\theta^4/(4\ln2)\approx1.05$\,bit/s/Hz at $25^\circ$ for the reweighting alone. The imaginary parts rotate each user's contribution. Section~\ref{sec:sim} shows that this rotation mainly slows the iteration down.
\end{remark}

\section{Estimating the Per-User Factors}\label{sec:e1}

The per-user factors are only $K$ numbers, and the array helps to estimate them. In the estimator E1, user $k$ alone transmits a known nonzero symbol $c$ over the uplink with the conjugation of Section~\ref{sec:model}, so that the BS obtains $v_k=c\,u_k(b\odot h_k)+z_k$. User $k$ reports the output $y_{kk}=h_k^Hw_k$, which it measures anyway during the forward probe. The BS forms
\begin{equation}
\hat u_k=\frac{w_k^Hv_k}{c\,\bar y_{kk}}
\end{equation}
and sends it back, and user $k$ then scales its seed by a capped version of $1/\hat u_k$, given below. The order in time is as follows. Uplink pilots give the initial precoder $W_0$, E1 runs once with $w_k$ taken from $W_0$, and the iterations follow. Without noise, $\hat u_k=u_k\beta_k$, where $\beta_k$ averages the BS factors with the complex weights of Lemma~\ref{lem:e1}. These weights can cancel, so $\hat u_k$ can be arbitrarily small even when $w_k^Hh_k\neq0$, and $1/\hat u_k$ is unbounded. For $h_k=(1,1)^\top$, $b=(1,2)^\top$, and $w_k\propto(2,-1)^\top$, for example, $\hat u_k=0$. Without a cap on the scale magnitude, the mean seed power is unbounded under Gaussian echo noise, because the noisy numerator of $\hat u_k$, given below, has a density that is positive at zero. User $k$ therefore transmits $\overline{q_ks_k}$ with
\begin{equation}\label{eq:cap}
q_k=e^{-j\arg\hat u_k}\min\bigl\{|\hat u_k|^{-1},\,2m,\,2\bigr\},
\end{equation}
where $m=\operatorname{median}_j|\hat u_j|^{-1}$.
The cap keeps each scale magnitude $|q_k|$ at most twice the median $m$, or $6$\,dB above it, and at most two, so no seed is transmitted more than $6$\,dB above its unscaled power. The relative cap triggers for $3$ to $10$\% of the users in the configurations of Fig.~\ref{fig:cal}. The fixed cap never triggered, since the largest $|q_k|^2$ over all simulated channels was $3.2$. The uplink noise is added after this scaling, so $q_k$ also sets the SNR of the returned field. At $W_0$, the scaled seeds of E1 carry $0.42$ to $0.97$ of the energy of the unscaled seeds, depending on the channel, and the returned field of E1 has $2.8$ to $3.9$\,dB lower SNR than without calibration. The one-time echo is not included. E1 therefore does not start from a stronger return than no calibration. The estimate $|\hat u_k/u_k|$ falls more than $10$\,dB below its channel median for $1.5$, $0.4$, and $0.1$\% of the users on the clustered channel at $M=64$, $128$, and $256$, and for at most $0.1$\% on the Rayleigh channel. With noise, the reported output is $y_{kk}+e_k$ and the echo carries $z_k$, so
\begin{equation}
\hat u_k=\frac{u_k\beta_kd_k+\eta_k}{d_k+\bar e_k},\qquad d_k=w_k^Hh_k,\ \eta_k=w_k^Hz_k/c,
\end{equation}
and at high SNR the error relative to the noiseless estimate $u_k\beta_k$ is $\eta_k/(u_k\beta_kd_k)-\bar e_k/d_k$ to first order. Both noises shrink in relative terms as the beam gain $|d_k|$ grows.

\begin{lemma}[E1]\label{lem:e1}
Let $w_k^Hh_k\neq0$, let the reported output $y_{kk}$ be exact, and let the returns be noiseless. Then $\hat u_k=u_k\beta_k$ with $\beta_k=\sum_ib_i\xi_{ik}/\sum_i\xi_{ik}$ and $\xi_{ik}=\bar w_{ik}h_{ik}$. If the $b_i$ are independent and identically distributed and independent of $(H,W)$, then conditionally on $(H,W)$, $\E\beta_k=\E b$ and
\begin{equation}
\operatorname{Var}\beta_k=\frac{\operatorname{Var}(b)}{N_k},\qquad N_k=\frac{|\sum_i\xi_{ik}|^2}{\sum_i|\xi_{ik}|^2}\le M.
\end{equation}
\end{lemma}
\begin{proof}
Substitute $v_k$ and $\bar y_{kk}=w_k^Hh_k=\sum_i\xi_{ik}$. The variance follows from the independence of the $b_i$, and $N_k\le M$ from the Cauchy--Schwarz inequality.
\end{proof}

The variance formula needs $W$ independent of $b$. When $W_0$ comes from pilots that pass through $D_bHD_u$, the beam depends on $b$. In the homogeneous case $|h_{ik}|=1$ with a matched beam $w_k\propto D_bh_k$, for instance, $\beta_k$ tends to $\E|b|^2/\overline{\E b}$, which is again common to all users and real and positive. Table~\ref{tab:e1} measures the realized error with the pilot-based initial precoder $W_0$. For this precoder the variance formula of Lemma~\ref{lem:e1} is not a prediction, and $N_k$ serves only as a geometric diagnostic.

$N_k$ is large when the products $\xi_{ik}$ are phase-aligned and of comparable magnitude across antennas, as for beams close to matched filtering. Spreading weight over the array is not enough by itself. The mean $\E b=e^{\sigma_{\ln}^2/2}e^{-\sigma_\theta^2/2}$ is real, positive, and common to all users, so gradient normalization removes it. The factors are drawn with zero-mean phase, and no arm removes or adds a common phase during the iterations. A finite array leaves a common phase of order $\sigma_\theta/\sqrt M$ in each realization, which is retained. The common amplitude and phase removal in Table~\ref{tab:e1} uses the true factors as a diagnostic only and is not applied in the runs of Fig.~\ref{fig:cal}. For a beam independent of $b$, the deviation $\beta_k-\E b$ has variance $\operatorname{Var}(b)/N_k$, depends on the complex beam weights, and differs from user to user. E1 costs $K$ uplink symbols, $K$ reported scalars, and $K$ scalars sent back, once per optimization run, from the $W_0$ of that run. A single iteration of physical backpropagation already uses $K$ downlink channel uses for the probe and $K$ uplink channel uses for the return, in which all users transmit simultaneously. With the $K$ probe and $K$ return symbols of an iteration placed on $K$ adjacent subcarriers of one downlink and one uplink orthogonal frequency-division multiplexing (OFDM) symbol, an iteration costs one TDD switching period. Equation~(1) then requires the same channel on these subcarriers. With $15$\,kHz spacing and $K=80$ they span $1.2$\,MHz, and we assume the channel is flat over this band. We count air time only. We assume a switching period of $0.5$\,ms, so $1000$ iterations take $0.5$\,s and $3000$ take $1.5$\,s. This is the regime of static users such as fixed wireless access, which we assume throughout. We assume that the hardware factors drift much more slowly than an optimization run, so one BS calibration serves many such runs, while E1 runs once per run.

\section{Simulations}\label{sec:sim}

\subsection{Setup}
$K=80$ users, $P=K$, and a nominal SNR $P/\sigma^2$ of $20$\,dB, so $\sigma^2=P/100$, unless stated otherwise. The Rayleigh channel has independent circularly symmetric complex Gaussian entries. The clustered channel uses a half-wavelength uniform linear array. It draws five cluster centers uniformly in $[-60^\circ,60^\circ]$ and assigns each user to a cluster uniformly at random. Each user's angle lies uniformly within $\pm5^\circ$ of its center, and its channel sums three to seven paths with angles within $\pm2.5^\circ$ of the user angle and independent complex Gaussian gains. Both channels are normalized to $\|H\|_F^2=MK$, so that the two models have equal average gain. The optimizer is that of Section~\ref{sec:model}, and all reported values are those of the last iterate after $3000$ iterations, from the single start $W_0$, unless stated otherwise.

Each user has a pilot of length $\tau=20$ with independent complex Gaussian entries, normalized to unit energy and collected in $\Phi\in\mathbb C^{\tau\times K}$. The BS receives $Y=\tilde H\Phi^\top+N_{\rm p}$ with noise variance $\sigma^2/P$ per entry, $10^{-2}$ in the baseline. Here $\tilde H=H$ for the reciprocal reference and $\tilde H=D_bHD_u$ for every mismatched arm. The BS forms the regularized estimate $\hat H=Y\bar\Phi(\Phi^\top\bar\Phi+\lambda_{\rm p}I)^{-1}$ with $\lambda_{\rm p}=100\sigma^2/P$, one in the baseline, and $\lambda_{\rm p}=10\sigma^2/P$ for $K=40$. The initial precoder is $W_0=(\hat H\hat H^H+(K\sigma^2/P)I)^{-1}\hat H$, scaled to power $P$. All mismatched arms, including both oracle arms and E1, share this $W_0$ built from the raw factors, and calibration acts only on the iterations. On the true channel $W_0$ attains $0.83$ to $0.87$ of the ideal initial sum rate on the clustered channel and $0.76$ to $0.81$ on the Rayleigh channel.

In the calibration experiments the forward outputs carry noise of variance $\sigma^2/12$ and the uplink carries noise of variance $\sigma^2/4$. E1 uses $c=1$, a report with noise of variance $\sigma^2/12$, and an echo with noise of variance $\sigma^2/4$ added after $D_b$, and it is computed once from $W_0$. Oracle calibration replaces the raw factor by a residual draw in the channel. It is not a digital multiplication by $D_b^{-1}$, so the receiver noise is unchanged. We write $b_{\rm res}$ and $u_{\rm res}$ for the residual draws. Every arm returns
\begin{equation}\label{eq:arms}
\hat G=D_{b'}HD_{u'}S+N
\end{equation}
with the same noise $N$. Here $(b',u')$ is $(b,u)$ without calibration, $(b_{\rm res},u)$ for oracle BS calibration, $(b,u_{\rm res})$ for oracle per-user calibration, $(b_1b_{\rm res},u)$ for BS calibration relative to the first antenna, and $(b,u\odot q)$ for E1. The factor amplitudes thus set the SNR of the returned field. At $W_0$ this SNR is $0$ to $0.9$\,dB lower with oracle per-user calibration than with oracle BS calibration, so per-user calibration does not gain from a higher SNR. The comparison therefore asks which side, repaired to the same residual accuracy from the same start, helps the iteration more. We call this the calibration priority. It does not include the training cost of each calibration. All noises are circularly symmetric complex Gaussian and independent across entries. The forward-probe and return noises are redrawn at every iteration and are the same for all arms, while the pilot and E1 noises are drawn once per channel. Each channel has one draw of $(b,u)$ with $3$\,dB and $25^\circ$ errors and one residual draw with $0.3$\,dB and $2^\circ$, shared by all arms, so the arms are paired. Each point of Fig.~\ref{fig:cal} averages $12$ channels, and a percent loss is the mean over channels of the per-channel ratio.

\subsection{Which Side to Calibrate}

\begin{figure*}[t]
\centering
\includegraphics[width=\textwidth]{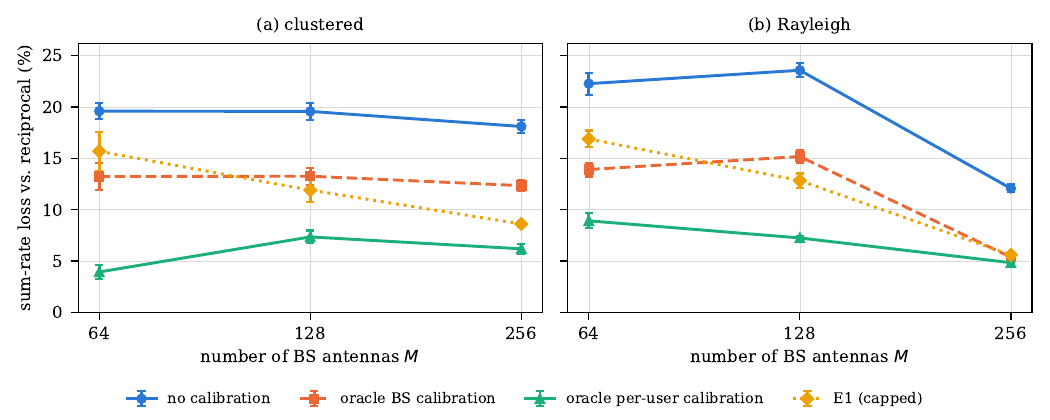}
\caption{Sum-rate loss relative to reciprocal physical backpropagation against the number of BS antennas after $3000$ iterations, with both sides mismatched by 3\,dB and $25^\circ$ and the initial precoder computed from mismatched pilots. Oracle calibration leaves a residual of 0.3\,dB and $2^\circ$. E1 uses the capped scale in \eqref{eq:cap}. All arms of a channel share one mismatch draw. Error bars show one standard error over 12 channels. The reciprocal reference reaches $158$, $261$, and $411$ bit/s/Hz on the clustered channel and $225$, $416$, and $589$ bit/s/Hz on the Rayleigh channel at $M=64$, $128$, and $256$.}
\label{fig:cal}
\end{figure*}

Fig.~\ref{fig:cal} compares four choices. On the clustered channel, oracle per-user calibration beats BS calibration at every tested array size, by paired differences of $9.3\pm1.5$, $5.9\pm0.7$, and $6.1\pm0.8$ percentage points at $M=64$, $128$, and $256$ (mean $\pm$ one standard error). At $M=256$ no calibration loses $18.1$\%, BS calibration $12.3$\%, and per-user calibration $6.2$\%. The BS calibration curve does not fall with $M$ on this channel, because the per-user amplitudes it leaves uncorrected keep changing who is served (Section~\ref{sec:mech}). On the Rayleigh channel the differences are $5.0\pm0.9$ and $7.9\pm0.7$ points at $64$ and $128$ antennas, and at $256$ antennas the two calibrations leave $5.4$ and $4.8$\%, a paired difference of $0.5\pm0.5$ percentage points, within one standard error. The ordering is wider at shorter budgets. At $1000$ iterations the paired differences are $11.8\pm1.8$, $9.4\pm0.8$, and $11.9\pm0.9$ points on the clustered channel and $2.4\pm1.3$, $10.1\pm1.1$, and $5.8\pm0.7$ points on the Rayleigh channel at $M=64$, $128$, and $256$. Only on the Rayleigh channel at $M=64$ below $500$ iterations does BS calibration lead. On six of these channels at $M=256$, running $10\,000$ iterations ($5$\,s of air time, a static diagnostic) changes the clustered difference from $7.2\pm1.2$ to $6.2\pm1.2$ points, with per-user calibration leaving $5.5$\% in both cases, and the Rayleigh difference stays within the error ($-0.8\pm0.6$). BS calibration relative to the first antenna leaves the common factor $b_1$ in \eqref{eq:arms}. Its phase rotates the whole returned field. Its amplitude scales the signal against the fixed noise and shifts the SNR by $20\log_{10}|b_1|$\,dB, which the normalization does not undo. In the six configurations of Fig.~\ref{fig:cal} this arm loses $1.0$ to $3.8$ points more than oracle BS calibration, and per-user calibration leads it in all six, by $4.1\pm0.7$ points on the Rayleigh channel at $M=256$. A control run splits $b_1$ into its two parts on the same channels. Keeping only the phase of $b_1$ costs $0.2$ to $3.7$ points more than oracle BS calibration, and keeping only its amplitude costs $-0.5$ to $1.0$ points, within $1.4$ standard errors of zero in every configuration. The common phase carries most of the extra loss. The amplitude shifts the SNR channel by channel and averages out.

\begin{table}[t]
\centering
\caption{E1 with the initial precoder from mismatched pilots, 12 channels with 80 users each (users of one channel are not independent). Standard deviation of $\hat u_k/u_k$ in amplitude (dB) and phase (deg), pooled over users and channels. As a diagnostic, the amplitudes in dB are centered at their per-channel median and the phases are measured from the angle of the per-channel mean of $\hat u_k/u_k$, using the true factors. The phase standard deviation is taken over wrapped angles. The row after the cap gives the effective residual $u_kq_k$, with the same normalization. The cap changes only the magnitude of $q_k$, so no phase row is repeated. The last row gives the per-channel median and 5th percentile of $N_k$, averaged over channels.}
\label{tab:e1}
\setlength{\tabcolsep}{4pt}
\begin{tabular}{lcccc}
\toprule
 & $M$ & 64 & 128 & 256 \\
\midrule
Clustered & amplitude & 3.97 & 2.53 & 1.21 \\
 & amplitude after cap & 3.42 & 2.10 & 1.05 \\
 & phase & 34.2 & 19.1 & 8.9 \\
 & $N_k$ median / 5\% & 3.7 / 0.34 & 11.8 / 2.0 & 31.3 / 6.8 \\
\midrule
Rayleigh & amplitude & 2.03 & 1.30 & 0.85 \\
 & amplitude after cap & 1.86 & 1.19 & 0.82 \\
 & phase & 16.2 & 8.7 & 5.3 \\
 & $N_k$ median / 5\% & 7.4 / 2.2 & 16.2 / 5.3 & 33.5 / 12.8 \\
\bottomrule
\end{tabular}
\end{table}

\begin{table}[t]
\centering
\caption{Sensitivity on the clustered channel with $M=128$ (8 channels, 12 for the baseline): loss in percent after 3000 iterations, and the paired difference between BS and per-user calibration. The baseline has $K=80$, SNR $20$\,dB, and $3$\,dB, $25^\circ$ errors on both sides. BS cal. and user cal. are oracle calibrations, E1 uses the capped scale, and $\pm$ gives one standard error.}
\label{tab:sens}
\begin{tabular}{lccccc}
\toprule
 & none & BS cal. & user cal. & E1 & BS$-$user \\
\midrule
Baseline & 19.6 & 13.3 & 7.4 & 11.9 & $5.9\pm0.7$ \\
SNR $10$\,dB & 19.0 & 8.4 & 12.0 & 42.8 & $-3.6\pm2.9$ \\
SNR $30$\,dB & 19.5 & 14.3 & 4.6 & 7.7 & $9.8\pm1.6$ \\
$K=40$ & 16.8 & 11.0 & 5.8 & 6.3 & $5.1\pm1.4$ \\
User $1$\,dB, $10^\circ$ & 8.5 & 2.0 & 7.2 & 11.4 & $-5.2\pm0.7$ \\
\bottomrule
\end{tabular}
\end{table}

Table~\ref{tab:sens} maps where the ordering holds. It holds at $30$\,dB and with $40$ users. It reverses when the user errors shrink to $1$\,dB and $10^\circ$, where the uncorrected BS chains dominate, and at $10$\,dB the two calibrations are within two standard errors of each other. With equal error strengths on the two sides, per-user calibration came first in every tested configuration except the Rayleigh channel at $M=256$ and the $10$\,dB case, where the two are within two standard errors. The priority therefore depends on the channel, the array size, and the SNR as well as on the relative error strength.

E1 approaches oracle per-user calibration as the array grows. Its paired excess over oracle per-user calibration is $11.7\pm2.0$, $4.6\pm0.9$, and $2.4\pm0.5$ points on the clustered channel at $M=64$, $128$, and $256$, and $8.0\pm0.6$, $5.6\pm0.6$, and $0.8\pm0.1$ points on the Rayleigh channel. Without the cap the excess is the same within $0.9$ points at $M=128$ and $256$, while at $M=64$ on the clustered channel uncapped E1 gives no gain over no calibration and the cap recovers $3.9$ points. Table~\ref{tab:e1} shows why. The effective number $N_k$ of Lemma~\ref{lem:e1} has a median of $3.7$ at $M=64$ on the clustered channel, and its 5th percentile falls below one. Table~\ref{tab:sens} adds two regimes where E1 should not be used. At $10$\,dB its errors reach $3.4$\,dB and $28^\circ$, larger than the mismatch it corrects, and it loses $43$\%. With $1$\,dB and $10^\circ$ user errors its own error of $2.0$\,dB and $14^\circ$ exceeds the user error. In the tested configurations E1 helped where its estimation error was below the user error and hurt where it was above. This is an observation, not a criterion. A smaller weight spread can still change the served users for the worse, as the two-user case of Proposition~\ref{prop:group} shows.

\subsection{Why the User Side Matters}\label{sec:mech}

\begin{figure}[t]
\centering
\includegraphics[width=\columnwidth]{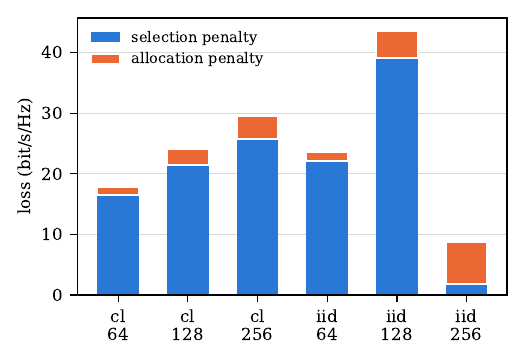}
\caption{The split of Proposition~\ref{prop:split} evaluated at the endpoints of the iteration alone, without WMMSE, for 3\,dB per-user amplitude errors, noise-free gradients, and a served-user threshold of 0.1\,bit/s/Hz, with ten starts and $10\,000$ iterations per objective. Each bar averages six channels, each with its own mismatch draw. ``cl'' denotes the clustered channel, ``iid'' the Rayleigh channel, and the number below is $M$.}
\label{fig:split}
\end{figure}

To examine the mechanism, we remove the measurement noise and track every user by identity. Each run keeps its best feasible iterate for its own objective, the start included. In this subsection the endpoint of a run is this iterate, and $W_1$ and $W_a$ denote the best computed solutions of the plain and weighted objectives. The starts are the pilot-based $W_0$ and perturbed copies $W_0+\epsilon\|W_0\|_FZ/\|Z\|_F$ with $Z$ of independent complex Gaussian entries and $\epsilon\in\{0.1,0.3\}$, and each objective keeps its best start. We write $\tilde F(\mathcal S)$ for the numerical estimate of $F(\mathcal S)$, obtained by re-optimizing the plain sum rate with the users outside $\mathcal S$ switched off, from $W_a$ and $W_1$ with those columns set to zero and perturbed copies of both. The plain optimum is estimated by the larger of $R(W_1)$ and $\tilde F(\mathcal S_1)$. Setting the columns to zero removes their interference, and rescaling to power $P$ raises every remaining SINR, so
\begin{equation}
F(\mathcal S_a)\ge R(W_a)-\sum_{k\notin\mathcal S_a}R_k(W_a).
\end{equation}
The computed allocation bracket can therefore be negative by at most the rate of the users below the threshold. This mass is below $10^{-4}$\,bit/s/Hz in every configuration, and no negative bracket occurred. The selection share is the ratio of the summed selection terms to the summed losses over the channels. Fig.~\ref{fig:split} shows the result.

The split is stable along the start ladder. Going from three starts and $3000$ iterations to ten starts and $10\,000$ iterations changes the selection share by at most one percentage point in every configuration. The step schedule leaves only $5$\% of its cumulative length after $3000$ iterations, so the budget extension is a weak test. An independent solver gives the stronger test. On the same channels we added the WMMSE algorithm~\cite{shi2011wmmse} to every optimization of the split, with $300$ iterations started from $W_0$ and from the best endpoint of the iteration. For each objective we kept the solution with the better value of that objective. The weighted endpoint $W_a$ is the better solution of the weighted objective, and its plain sum rate and served set $\mathcal S_a$ are computed on the same matrix. Each masked problem is solved by the iteration from $W_a$ and $W_1$, with the columns outside the mask set to zero and the power rescaled to $P$, and by WMMSE from the masked $W_0$ and from both of these solutions. The iteration used three starts of $3000$ iterations. The selection share becomes $90$, $86$, and $88$\% on the clustered channel and $94$, $90$, and $21$\% on the Rayleigh channel at $M=64$, $128$, and $256$, against $93$, $90$, $87$, $94$, $90$, and $21$\% with the iteration alone. WMMSE raised the weighted objective by $0$ to $4.6$\,bit/s/Hz on average per configuration. It changed the selection term by $-4.7$ to $+1.4$\,bit/s/Hz and the allocation term by at most $0.3$\,bit/s/Hz, so the two solvers differ mainly in which users the weighted endpoint serves. WMMSE raised the plain sum rate by $1.3$ to $3.0$\,bit/s/Hz on the clustered channel and on the Rayleigh channel at $M=64$. On the Rayleigh channel at $M=128$ and $256$ it improved no endpoint of the iteration, and in a spot check the endpoint was a fixed point of the WMMSE update. In an earlier protocol with one start per objective, a sweep over thresholds of $0.01$, $0.1$, and $0.3$\,bit/s/Hz left the selection share unchanged to the shown precision. Local solvers can still miss optima, so we also checked the protocol on an instance with an exact solution: eight orthogonal groups of two users with collinear channels and no group-power constraint, where every optimum serves at most one user per group (Proposition~\ref{prop:group}) and enumeration of the $2^8$ selections gives the exact split. Over $50$ draws the protocol recovered the plain optimum exactly. The weighted ascent, however, often stayed at a local maximum near the plain served set, and the computed selection term never exceeded the exact one. In this control, solver error therefore understates the selection penalty rather than inflating it.

Near full service, the amplitude-induced loss is close to the value of Proposition~\ref{prop:fullservice}. On the Rayleigh channel at $M=256$, the plain endpoints serve all 80 users and the weighted ones $79.7$ on average. The loss is $8.8$\,bit/s/Hz or $1.40$\%, its allocation part is $6.9$\,bit/s/Hz, and Proposition~\ref{prop:fullservice} gives $7.0$\,bit/s/Hz for the same weights. This is a numerical closeness, not a test of Proposition~\ref{prop:fullservice}, whose zero-forcing and full-activity assumptions these endpoints do not satisfy exactly, and the selection term is still $1.9$\,bit/s/Hz there, about $21$\%. In the other five configurations the endpoints leave many users unserved, and the selection term carries $87$ to $93$\% of the loss, and each configuration's $95$\% bootstrap interval over channels lies within $0.84$ to $0.95$. The turnover is much larger than the change in the number of served users. On the Rayleigh channel at $M=64$, $20.0$ users drop out and $12.7$ enter on average, a net change of $7.3$. At $M=256$ the clustered plain endpoints still leave $23.2$ of 80 users unserved, so the amplitude errors keep paying the selection penalty there.

Phase errors act differently. On the Rayleigh channel at $M=256$ with only $25^\circ$ per-user phase errors, the loss is $24.8$\,bit/s/Hz after $3000$ iterations, $2.0$\,bit/s/Hz after $6000$, and $1.7$\,bit/s/Hz or $0.27$\% after $10\,000$. On the clustered channel it falls from $24.4$ to $6.7$\,bit/s/Hz, or $1.64$\%. These are endpoints of a finite budget, not stationary points. The sum rate changes by less than $0.01$\,bit/s/Hz between $8000$ and $10\,000$ iterations only because the step has decayed, and the component of the returned field tangent to the power sphere is still $0.33$ to $0.51$ of its norm. Restricted to zero-forcing directions with a power-only update, the same errors cost $1.62$\,bit/s/Hz on the Rayleigh channel, which equals the cosine-weight value $K\log_2(A/G)$ of Remark~1 for these draws. On the Rayleigh channel the loss after $10\,000$ iterations thus has the size of the cosine-weight term. On the clustered channel the zero-forcing control loses only $1.0$\,bit/s/Hz, so most of the remaining $6.7$\,bit/s/Hz is not explained by the cosine weights, and we leave its mechanism open. At a budget of a few thousand iterations the main cost of the phase errors is slow progress, so per-user calibration has to correct the phase as well, and E1 does so because $\hat u_k$ is complex.

With only the BS errors, the loss after $3000$ iterations is $7.4$ and $11.5$\,bit/s/Hz on the clustered and Rayleigh channels at $M=64$, and $22.8$ and $26.7$\,bit/s/Hz at $M=256$, between $4.1$ and $5.0$\% in all four cases. This is below the $31.4$ and $35.7$\,bit/s/Hz of the $K$-user extrapolation of Proposition~\ref{prop:single} for the same draws, and close to the $4.8$ to $6.2$\% that per-user calibration leaves in Fig.~\ref{fig:cal} at $M=256$. After BS calibration, the uncorrected amplitude errors still change who is served, and the phase errors still slow the iteration.

\section{Conclusion}\label{sec:concl}

In physical backpropagation with ideal BS chains and noiseless forward outputs and returns, positive per-user factors change the objective, while the BS chain factors scale the returned field antenna by antenna. Per-user amplitudes change who is served, and per-user phases slow the iteration. For one user the BS chains leave an SNR loss with a closed form that the array does not remove. With 80 users the BS-only loss is $4.1$ to $5.0$\% in our simulations. With pilots through the mismatched chains and paired hardware draws, oracle per-user calibration beats BS calibration on clustered channels at every tested array size and on Rayleigh channels up to $128$ antennas, and the ordering reverses when the user errors are much smaller than the BS errors. The single-echo estimator E1 came within $2.4$ percentage points of oracle per-user calibration at $256$ antennas and hurt in the tested regimes where its own error exceeded the user error.

These results come from simulations with static mismatch with lognormal amplitudes and Gaussian phases, evaluated after a fixed iteration budget from one pilot-based start, and from a few starts in the mechanism study. Channels that change within an optimization run and a closed form for the multiuser BS-side loss are left open.

\bibliographystyle{IEEEtran}
\bibliography{refs}

\end{document}